\UseRawInputEncoding
\documentclass[twocolumn,prb,notitlepage,floats,citeautoscript,superscriptaddress,amsmath,amssymb,utf8]{revtex4-2}

\usepackage[version=3]{mhchem} 
\usepackage{bm}
\usepackage{inputenc}
\usepackage[T1]{fontenc}
\usepackage{graphicx}
\usepackage{upgreek}
\usepackage{color}
\usepackage[normalem]{ulem}
\usepackage{hyperref} 
\hypersetup{colorlinks,citecolor=blue, filecolor=blue ,linkcolor=blue , urlcolor=blue, pdftex}
\usepackage{sidecap}
\usepackage{amssymb}
\usepackage[caption=false]{subfig}
\usepackage{multirow}

\def \FUW{Institute of Experimental Physics, Faculty of Physics, University of Warsaw, 02-093 Warsaw, Poland}
\def \Watanabe{Research Center for Electronic and Optical Materials, National Institute for Materials Science, 1-1 Namiki, Tsukuba 305-0044, Japan}
\def \Taniguchi{Research Center for Materials Nanoarchitectonics, National Institute for Materials Science, 1-1 Namiki, Tsukuba 305-0044, Japan}
\def \Grenoble {Laboratoire National des Champs Magn\'etiques Intenses, CNRS-UGA-UPS-INSA-EMFL, 25 avenue des Martrys, 38042 Grenoble, France}
\def \Centera{CENTERA Laboratories, Institute of High Pressure Physics, Polish Academy of Sciences, 01-142 Warsaw, Poland}
\def \Singapore{Institute for Functional Intelligent Materials, National University of Singapore, 117544, Singapore}

\begin{document}

\title{The temperature influence on the brightening of neutral and charged dark excitons in WSe$_2$ monolayer}

\author{\L{}ucja Kipczak}
\email{lucja.kipczak@fuw.edu.pl}
\affiliation{\FUW}
\author{Natalia Zawadzka}
\affiliation{\FUW}
\author{Dipankar Jana}
\affiliation{\Grenoble}
\affiliation{\Singapore}
\author{Igor Antoniazzi}
\affiliation{\FUW}
\author{Magdalena~Grzeszczyk}
\affiliation{\Singapore}
\author{Ma\l{}gorzata Zinkiewicz}
\affiliation{\FUW}
\author{Kenji~Watanabe}
\affiliation{\Watanabe}
\author{Takashi Taniguchi}
\affiliation{\Taniguchi}
\author{Marek~Potemski}
\affiliation{\FUW}
\affiliation{\Grenoble}
\affiliation{\Centera}
\author{Cl\'ement Faugeras}
\affiliation{\Grenoble}
\author{Adam Babi\'nski}
\affiliation{\FUW}
\author{Maciej R. Molas}
\email{maciej.molas@fuw.edu.pl}
\affiliation{\FUW}

\begin{abstract}
The optically dark states play an important role in the electronic and optical properties of monolayers (MLs) of semiconducting transition metal dichalcogenides.
The effect of temperature on the in-plane-field activation of the neutral and charged dark excitons is investigated in a WSe$_2$ ML encapsulated in hexagonal BN flakes.
The brightening rates of the neutral dark (X$^\textrm{D}$) and grey (X$^\textrm{G}$) excitons and the negative dark trion (T$^\textrm{D}$) differ substantially at a particular temperature. More importantly, they vanish considerably by about 3 -- 4 orders of magnitude with the temperature increased from 4.2~K to 100~K.
The quenching of the dark-related emissions is accompanied by the two-order-of-magnitude increase in the emissions of their neutral bright counterparts, $i.e.$ neutral bright exciton (X$^\textrm{B}$) and spin-singlet (T$^\textrm{S}$) and spin-triplet  (T$^\textrm{T}$) negative trions, due to the thermal activations of dark states.
Furthermore, the energy splittings between the dark X$^\textrm{D}$ and T$^\textrm{D}$ complexes and the corresponding bright X$^\textrm{B}$, T$^\textrm{S}$, and T$^\textrm{T}$ ones vary with temperature rises from 4.2~K to 100~K.
This can be explained in terms of the different exciton-phonon couplings for the bright and dark excitons stemming from their distinct symmetry properties.
\end{abstract}

\maketitle

\section{Introduction \label{sec:Intro}}
The existence of dark neutral and charged (trions) excitons determines the optical response in the so-called $darkish$ monolayers (MLs) of semiconducting transition metal dichalcogenides (S-TMDs), $i.e.$ MoS$_2$, WS$_2$, and WSe$_2$~\cite{Molas2017, Zhang2017, Robert2017, Molas2019, Liu2019, LiuValley, Lu2020, Robert2020, Znkiewicz2020nanoscale, Arora2020, Liu2020, He2020,  Zinkiewicz2021nanolett, Kapuściński2021, Zinkiewicz2022}.
For $darkish$ MLs, the dark excitons and trions are characterised by significantly lower energies compared to their bright counterparts~\cite{Molas2017, Zhang2017, Robert2017, Molas2019, Liu2019, LiuValley, Lu2020, Robert2020, Znkiewicz2020nanoscale, Arora2020, Liu2020, He2020,  Zinkiewicz2021nanolett, Kapuściński2021, Zinkiewicz2022}.
Alternatively, the $bright$ S-TMD MLs, $i.e.$ MoSe$_2$ and MoTe$_2$, have the energetically lowest states, which are optically allowed. 
The dark transitions are optically forbidden or inactive, as the recombining electron-hole ($e$-$h$) pairs are characterised by a parallel spin configuration for an electron and a hole of the conduction and valence bands, respectively~\cite{Koperski2017, Molas2017}. 
To date, the properties of dark excitons and trions have been investigated in MLs embedded between different environments ($e.g.$ exfoliated on Si/SiO$_2$ substrate or encapsulated in hexagonal boron nitride (hBN) flakes)~\cite{Molas2017, Zinkiewicz2022}, for various levels of free carrier concentration~\cite{Liu2019, Liu2020, He2020}, or due to the Lamb shift~\cite{Ren2023}.
The most well-known phenomenon for significantly brightening of the emission due to dark states in S-TMD MLs is achieved by the in-plane magnetic field ($B_\parallel$), which leads to mixing the spin levels of bright and dark excitons~\cite{Molas2017, Zhang2017, Lu2020, Robert2020}.

In this work, we determine the temperature influence on the in-plane-field activation (brightening) of neutral dark (X$^\textrm{D}$) and grey (X$^\textrm{G}$) excitons and negative dark trions (T$^\textrm{D}$) in a high-quality WSe$_2$ ML encapsulated in hBN flakes.
We found that the brightening of the X$^\textrm{D}$, X$^\textrm{G}$, and T$^\textrm{D}$ lines differ substantially from each other at a particular temperature, but more importantly, it vanishes considerably by almost 3 -- 4 orders of magnitude with the temperature increased from 4.2~K to 100~K.
The quenching of the dark-related emissions is accompanied by the two-order-of-magnitude enlargement of the neutral bright counterparts, $i.e.$ neutral bright exciton (X$^\textrm{B}$) and spin-singlet (T$^\textrm{S}$) and spin-triplet  (T$^\textrm{T}$) negative trions, due to the thermal activations of bright states.
In addition, the extracted dark-bright energy splitting between the neutral and charged complexes is also affected when the temperature is increased from 4.2~K to 100~K.
This can be explained in terms of the different exciton-phonon couplings for the bright and dark excitons because of their various symmetries.

\section{Experimental Results \label{sec:results}}

\begin{figure*}[!th]
	\subfloat{}%
	\centering
	\includegraphics[width=0.95 \linewidth]{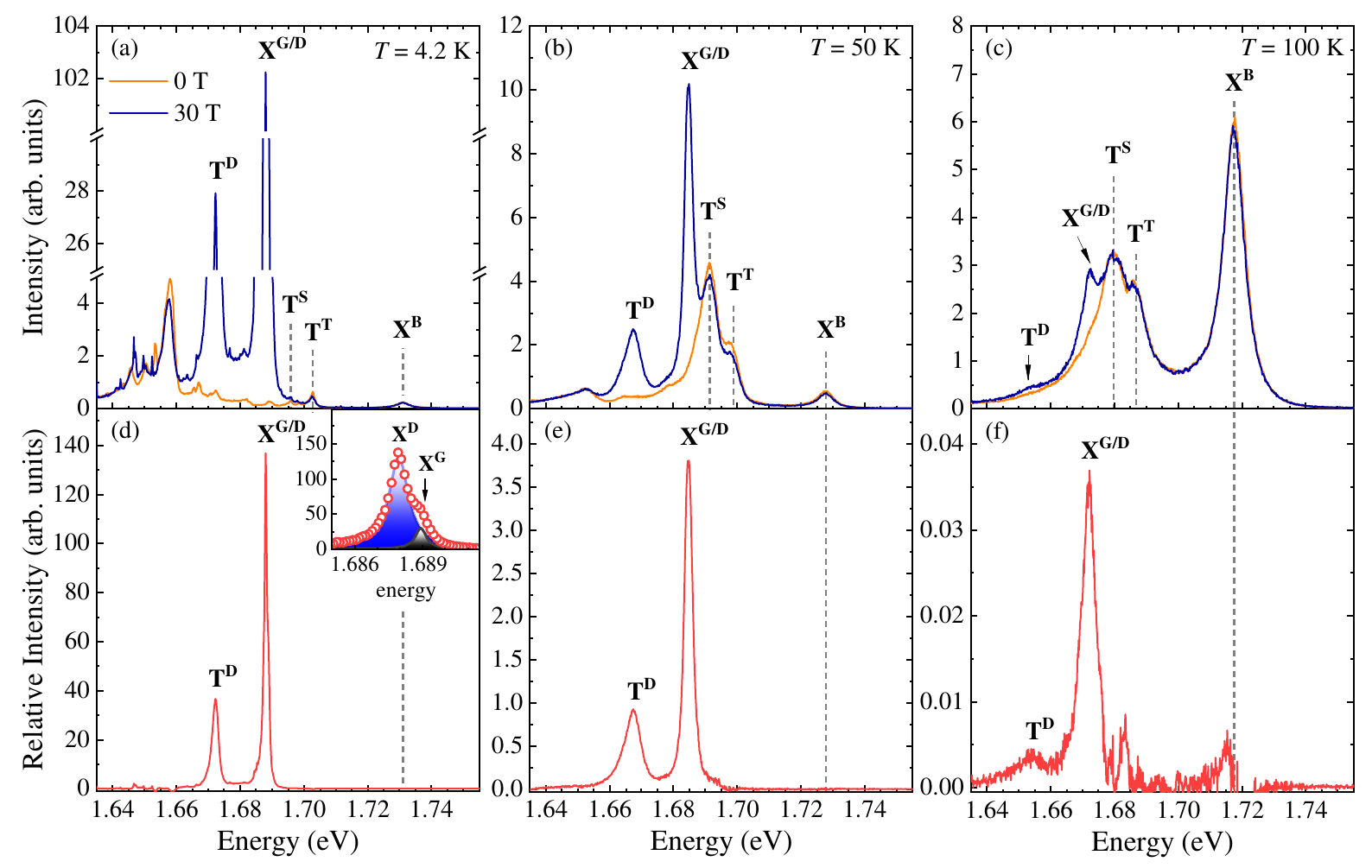}
   	\caption{(a)-(c) PL spectra of the investigated WSe$_2$ ML at different temperatures: (a) 4.2~T, (b) 50~K, and (c) 100~K, measured at zero field (orange curves) and at $B_\parallel$=30~T (blue curves) applied in the plane of the ML. The PL spectra were normalised to the intensity of the bright X$^\textrm{B}$ line.(d)-(f) Corresponding relative spectra (red curves) deﬁned as $\textrm{PL}_{B_\parallel=30~\textrm{T}} - \textrm{PL}_{B_\parallel=0~\textrm{T}}$.}
    \label{fig:1}
\end{figure*}

Figures~\ref{fig:1}(a)-(c) represent the photoluminescence (PL) spectra of the WSe$_\textrm{2}$ ML encapsulated in hBN flakes at zero magnetic field and at 30~T applied in the plane of the ML ($B_\parallel$) measured at three different temperatures 4.2~K, 50~K, and 100~K, respectively. 
The zero-field spectrum at 4.2~K displays a set of characteristic emission lines; see Fig.~\ref{fig:1}(a).
This spectrum, apart from those related to the neutral bright exciton (X$^\textrm{B}$), intravalley spin-singlet (T$^\textrm{S}$) and intervalley spin-triplet (T$^\textrm{T}$) negative trions, consists of several emission lines on its lower energy side.
These additional lines have been attributed in the literature to charged excitons (trions), neutral and charged biexcitons, dark excitons and trions, their phonon replicas, and etc.~\cite{Courtade2017, Li2018, Barbone2018, Chen2018, Paur2019, Li2019, Liu2019, Li2019momentum, LiuValley, Liu2020, Molas2019, Arora2020, Liu2020, He2020,  Zinkiewicz2021nanolett, Kapuściński2021, Zinkiewicz2022}, described in detail in the Supplementary Material (SM). 
The application of the external $B_\parallel$ field results in the appearance of the new signal at energies below the X$^\textrm{B}$, see Figs.~\ref{fig:1}(a)-(c).
These new field-induced emission lines were associated with the recombination processes of neutral dark exciton (X$^\textrm{D}$) and negative dark trion (T$^\textrm{D}$).
The sign of the T$^\textrm{D}$ trion is determined from the PL lineshape at $T$=4.2~K in which two lines due to negative T$^\textrm{S}$ and T$^\textrm{T}$ trions are apparent~\cite{Courtade2017}.

To better visualise the effect of the in-plane magnetic field on dark complexes and compare the results obtained at different temperatures, we define the relative spectrum as $\textrm{PL}_{B_\parallel=30~\textrm{T}} - \textrm{PL}_{B_\parallel=0~\textrm{T}}$ intensity. 
The relative spectra obtained for $B_\parallel$=30~T are shown in Fig.~\ref{fig:1}(d)-(f).
For dark excitons, the exchange interaction lifts their valley degeneracy, which gives rise to a fine structure splitting with two types of states, termed grey and dark excitons~\cite{Slobodeniuk2016, Robert2017, Molas2019}. 
These two states are qualitatively different.
The X$^\textrm{G}$ has an optically active recombination channel with photons emitted within the ML plane~\cite{Wang2017}, which can be observed only in a standard out-of-plane experimental setup when using objectives with a high numerical aperture~\cite{Robert2017, Molas2019}.
The X$^\textrm{D}$ state is truly optically forbidden, and its activation requires external magnetic fields~\cite{Robert2017, Molas2019}.
The detailed analysis of the emission line, shown in the inset to Fig.~\ref{fig:1}(d), reveals its fine structure, $i.e.$ X$^\textrm{D}$ and X$^\textrm{G}$, which is in line with previous studies on WSe$_2$ MLs~\cite{Robert2017, Molas2019}.
The dark exciton emission was deconvoluted using two Lorentz functions up to $T$=40~K, which allowed us to independently investigate the intensities of the contributions of the X$^\textrm{G}$ and X$^\textrm{D}$ transitions.

As seen in Fig.~\ref{fig:1}, three lines, $i.e.$ T$^\textrm{D}$, X$^\textrm{D}$, and X$^\textrm{G}$, significantly brightened at $T$=4.2~K.
The temperature increase to 50~K results in the forty and thirty five times decrease of the maximum intensity of the T$^\textrm{D}$ and X$^\textrm{D}$ emissions, respectively.
The further increase of temperature results in the intensity reduction by more than an order of magnitude for both emission lines.
This demonstrates that temperature can drastically modify the brightening effect of in-plane magnetic fields on the dark states.

\begin{figure}[t!]
		\subfloat{}%
		\centering
		\includegraphics[width=0.81\linewidth]{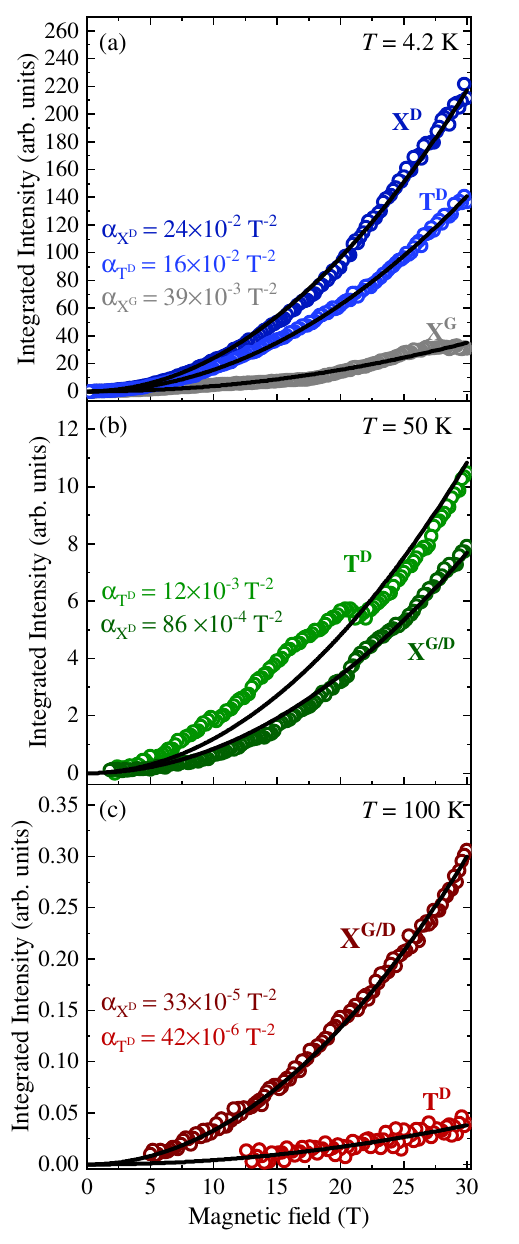}
    	\caption{Magnetic field evolution of the integrated intensities of the neutral grey (X$^\textrm{G}$) and dark (X$^\textrm{D}$) excitons and the dark trion (T$^\textrm{D}$) extracted at different temperatures: (a) 4.2~K, (b) 50~K, and (c) 100~K. Black curves represent the fits of the function $I = \alpha B_\parallel^{2}$.}
		\label{fig:2}
\end{figure}

\begin{figure}[b!]
		\subfloat{}%
		\centering
		\includegraphics[width=0.75\linewidth]{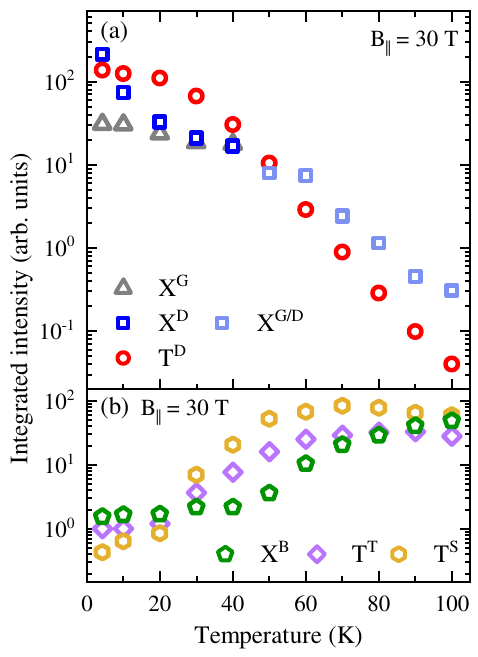}
    	\caption{ The integrated intensities of the (a) dark (X$^\textrm{G}$, X$^\textrm{D}$, and T$^\textrm{D}$) and (b) bright (X$^\textrm{B}$, T$^\textrm{T}$, and T$^\textrm{S}$) complexes, obtained in magnetic field $B_\parallel$=30~T, as a function of temperature. The vertical axes are given in logarithmic scale for clarity.}
		\label{fig:3}
\end{figure}

To analyse in detail the influence of temperature on the brightening of dark complexes, we deconvoluted the relative spectra using Lorentz functions.
The extracted evolutions of the integrated intensities ($I$) of the studied T$^\textrm{D}$, X$^\textrm{D}$, and X$^\textrm{G}$ complexes as a function of the in-plane magnetic field for three selected temperatures are presented in Fig.~\ref{fig:2}.
The dependencies are expected to be quadratic and can be described by the formula: $I = \alpha B^{2}_{\parallel}$, where $\alpha$ corresponds to the brightening coefficient (see Refs.~\cite{Slobodeniuk2016, Zhang2017, Molas2017, Molas2019}).
The $\alpha$ parameter is proportional to the population of dark excitons and the emission intensity of the bright exciton~\cite{Molas2019, Zinkiewicz2022}.
As can be seen in Fig.~\ref{fig:2}, the experimental data can be nicely reproduced by the fitted intensity evolutions given by the formula.
The complete analysis of the magnetic field dependences of the integrated intensities of the investigated lines are given in the SM.
The $\alpha$ is substantially different for the X$^\textrm{D}$ and X$^\textrm{G}$ excitons at $T$=4.2~K, $i.e.$ $\alpha_{\textrm{X}^\textrm{D}}$=24$\times$10$^{-2}$~T$^{-2}$ and $\alpha_{\textrm{X}^\textrm{G}}$=39$\times$10$^{-3}$~T$^{-2}$.
This large variation can be explained by the difference in the populations of these two states at 4.2~K, the population ratio $e^{-\delta/{kT}}$=0.161 ($\delta$=660~$\mu$eV~\cite{Molas2019}) is in excellent agreement with the measured ratio $\alpha_{\textrm{X}^\textrm{G}}/\alpha_{\textrm{X}^\textrm{D}}$=0.161.
This suggests that the relative population of the neutral grey and dark excitons is controlled by the Boltzmann distribution.
The $\alpha_{\textrm{T}^\textrm{D}}$=16$\times$10$^{-2}$~T$^{-2}$ confirms that the free-electron density in the studied WSe$_2$ ML is not negligible, leading to the formation of a significant number of dark trions.
From $T$=50~K, a single emission line can be resolved in the energy range of the neutral grey and dark excitons, which is denoted as X$^\textrm{G/D}$ in the following (see Fig.~\ref{fig:3}).
By analysing the temperature variation of the $\alpha$ parameters of the X$^\textrm{G}$ and X$^\textrm{D}$ lines, we can propose the attribution of the aforementioned single line at temperatures higher than 40~K.
Although the $\alpha_{\textrm{X}^\textrm{D}}$ parameter decreases by about one order of magnitude from 4.2~K to 50~K and simultaneously the variation of $\alpha_{\textrm{X}^\textrm{G}}$ is only a few times, we tentatively ascribe this line to the grey exciton.
The neutral and charged dark exciton intensities also follow the aforementioned quadratic evolutions at higher temperatures, see Figs.~\ref{fig:2}(b)-(c) and the SM for details.
However, the magnitude of magnetic activation of dark complexes, $i.e.$ $\alpha$ parameter, is greatly reduced by more than 3 orders of magnitude with increasing temperature from 4.2~K to 100~K.

The temperature dependence of the integrated intensities of the T$^\textrm{D}$, X$^\textrm{D}$, and X$^\textrm{G}$ complexes obtained in magnetic field $B_\parallel$=30~T is shown in Fig.~\ref{fig:3}(a). 
Note that the corresponding evolution of the extracted $\alpha$ parameters displays analogous trends and is shown in the SM.
The X$^\textrm{D}$ intensity reduces dramatically 7 times in the temperature range 4.2~K -- 20~K, while the X$^\textrm{G}$ counterpart stays almost at the same level.
At higher temperatures, the intensities of both neutral dark excitons show similar intensities up to 40~K, which is followed by the exponential decay of the X$^\textrm{D}$ intensity.
For the T$^\textrm{D}$, its intensity stays at the same level up to 20~K, and then a significant drop of T$^\textrm{D}$ is apparent.
In summary, the increase in temperature from 4.2~K to 100~K leads to the reduction of the integrated intensities of the dark complexes by more than 3 orders of magnitude. 
Figure~\ref{fig:3}(b) shows the integrated intensities of the X$^\textrm{B}$, T$^\textrm{T}$, and T$^\textrm{S}$ lines.
The evolution of the X$^\textrm{B}$ intensity describes a similar evolution to the X$^\textrm{G}$ and X$^\textrm{B}$ ones, which are, however, inverted.
This means that up to around 40 K - 50~K, a small increase of the X$^\textrm{B}$ intensity is observed, which is followed by its rapid exponential growth, as previously reported~\cite{Zhang2015}.
The analogous inverted evolution can be seen for the T$^\textrm{T}$ and T$^\textrm{S}$ lines compared to the dark trion.
These results show that the intensities of the bright and neutral complexes for a given family (neutral and charged) are associated with each other.
However, for the bright complexes, only a 2 orders of magnitude increase is observed in the range from 4.2~K to 100~K, as compared to the aforementioned 3 orders of magnitude reduction for the dark features.

Although the difference in the $\alpha$ parameter of the X$^\textrm{G}$ and X$^\textrm{D}$ lines can be explained by the difference in the population of these two states (the Boltzmann distribution) from 4.2~K to 40~K, the difference in the temperature evolutions between the dark and bright excitons and trions is more complex.
For bright neutral excitons in WSe$_2$ ML, strong quenching of their emission was reported to be associated with the presence of their dark counterparts with decreasing temperature~\cite{Arora2015, Zhang2015}.
In our case, we observe directly that the reductions mentioned above in the X$^\textrm{B}$, T$^\textrm{T}$, and T$^\textrm{S}$ emissions occur simultaneously with substantial brightenings of the dark complexes (X$^\textrm{D}$, X$^\textrm{G}$, and T$^\textrm{D}$).
Due to the substantial suppression of the thermal activation of dark excitons at low temperatures ($T<50$~K and $<30$~K for the neutral and charged excitons, respectively), bright emissions are also reduced in this temperature range.
When the temperature is increased, the efficiency of thermal activation from the dark to bright states increases, which leads to a shrinkage of the brightening of dark states as the population of the bright states grows.
The competition between the thermal activation of the bright states and the in-plane magnetic activation of the dark states suppresses the brightening of dark complexes at temperatures higher than 100~K.
An analogous quenching of the brightening, $i.e.$ above 100~K, was reported for WSe$_2$ ML exfoliated on the Si/SiO$_2$ substrate~\cite{Zhang2017}.

\begin{figure}[t!]
		\subfloat{}%
		\centering
		\includegraphics[width=0.69\linewidth]{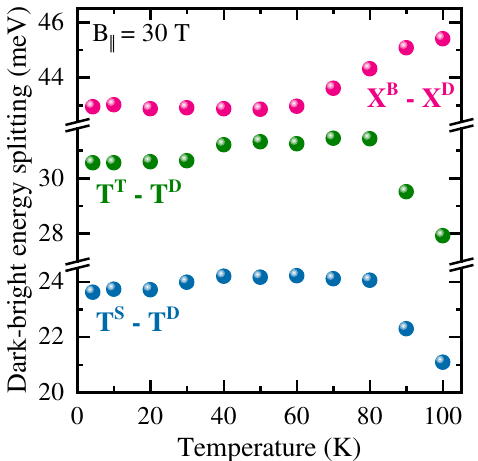}
    	\caption{The temperature evolution of the energy splitting between bright (X$^\textrm{B}$, T$^\textrm{T}$, and T$^\textrm{S}$) and dark (X$^\textrm{D}$, X$^\textrm{G}$, and T$^\textrm{D}$) complexes extracted from data measured in $B_{\parallel}$=30~T.}
		\label{fig:4}
\end{figure}

Figure~\ref{fig:4} presents the energy splitting between bright (X$^\textrm{B}$, T$^\textrm{T}$, and T$^\textrm{S}$) and dark (X$^\textrm{D}$, X$^\textrm{G}$, and T$^\textrm{D}$) complexes as a function of temperature.
The temperature evolutions of given excitonic complexes are shown in the SM.
The neutral bright-dark splitting (X$^\textrm{B}$ and X$^\textrm{D}$) stays constant at 60~K, and further a blueshift of about 3~meV is observed at higher temperatures. 
This behaviour can be understood in terms of different properties of the symmetries of the bright and grey (dark) excitons.
The X$^\textrm{B}$ exciton is characterised by the in-plane dipole momentum, whereas the zero and out-of-plane dipole momenta determine the X$^\textrm{D}$ and X$^\textrm{G}$ complexes. 
The typical redshifts of excitonic resonances under increased temperature are associated with the shrinkage of the bandgap energy~\cite{Unlu1992}. 
However, for excitonic complexes, the interaction of bound electron-hole pairs with lattice phonons needs to be taken into account, which results in slightly different temperature evolutions of particular excitonic transitions~\cite{Arora2015, Molas2017}.
Due to the different orientations of excitonic dipole moments, they can couple to phonons of several symmetries, $e.g.$ described by in-plane or out-of-plane vibrations, and hence affect the excitonic temperature dependences.
The bright-dark splittings for the negative trions (T$^\textrm{T}$ and T$^\textrm{S}$ versus T$^\textrm{D}$) are almost constant up to 80~K, and are followed by redshifts of about 2~meV.
In this case, unfortunately, the extracted dependences are a consequence of a peculiar temperature evolution of the T$^\textrm{D}$ lines (see the SM for details), which needs to be further developed.
The presented analysis of the bright-dark splitting sheds new light on the temperature activation of the bright complexes.
So far, its modelling relies on a constant value of energy splitting~\cite{Zhang2015, Arora2020}, while our results demonstrate its clear variation at higher temperature.
We believe that our work would trigger more theoretical studies on this issue.

\section{Summary \label{sec:Summary}}

In conclusion, we described the in-plane-field optical activation of the neutral dark/grey excitons and the negative dark trions in a WSe$_2$ ML as a function of temperature from 4.2~K to 100~K.
The brightening ratios of the dark complexes differ substantially from each other at a particular temperature, but more importantly, it vanishes considerably by about 3 -- 4 orders of magnitude with the temperature increased from 4.2~K to 100~K.
The quenching of the dark-related emissions is found to be accompanied by enlargement of the neutral bright counterparts, neutral bright exciton and spin-singlet and spin-triplet negative trions, due to their thermal activations.
Furthermore, the extracted dark-bright energy splittings between the neutral and charged complexes were shown to be a function of temperature with nonmonotonic changes of about 2~meV when temperature is increase from 4.2~K to 100~K.
This was explained in terms of the different exciton-phonon couplings for the bright and dark excitons because of their various symmetry properties.
Our results indicate that the population of dark excitons and dark trions plays a very important role in the WSe$_2$ ML emission spectra in the temperature range below 100~K, but also affect the spectra at higher temperatures due to increased thermal activation.

\section{Methods \label{sec:methods}}
The investigated sample is composed of a WSe$_2$ ML and hBN layers that were fabricated by two-stage PDMS (polydimethylsiloxane)-based mechanical exfoliation~\cite{Gomez2014}.
The WSe$_2$ ML flake was placed on the thick bottom hBN flake, which was directly exfoliated on the SiO2(90 nm)/Si substrate. 
Finally, the ML was capped with a thin top hBN flake and then the hBN/WSe$_2$ ML/hBN/SiO$_2$/Si structure was obtained.

Low-temperature micro-magneto-PL experiments were performed in the Voigt geometry, $i.e.$ magnetic field orientated parallel with respect to ML's plane. 
Measurements (spatial resolution ~1 $\mu$m) were carried out with the aid of a resistive magnetic coil that produces fields up to 30 T using a free-beam-optics arrangement. 
The sample was placed on top of a x-y-z piezo-stage kept at $T$=4.2~K and was excited
using a CW laser diode with 515 nm wavelength (2.41 eV photon energy). 
The emitted light was dispersed with a 0.5 m focal length monochromator and detected with a CCD camera.

\section{Acknowledgments \label{sec:Acknowledgments}}
We are grateful to Artur Slobodeniuk for fruitful discussions.
The work has been supported by the National Science Centre, Poland (Grant No. 2018/31/B/ST3/02111) and the CNRS via IRP '2DM' project. 
We acknowledge the support of the LNCMI-CNRS, member of the European Magnetic Field Laboratory (EMFL).
The Polish participation in EMFL was supported by the DIR/WK/2018/07 Grant from the Polish Ministry of Education and Science. 
K.W. and T.T. acknowledge support from the JSPS KAKENHI (Grant Numbers 21H05233 and
23H02052) and the World Premier International Research Center Initiative (WPI), MEXT,
Japan. 
M.P. acknowledges the support from the Foundation for Polish Science (MAB/2018/9 Grant within the IRA Program ﬁnanced by EU within SG OP Program).

\bibliographystyle{apsrev4-2}
\bibliography{biblio}

\end{document}


\title{Supplementary Material \\ The temperature influence on the brightening of neutral and charged dark excitons in WSe$_2$ monolayer}

\author{\L{}ucja Kipczak}
\email{lucja.kipczak@fuw.edu.pl}
\affiliation{\FUW}
\author{Natalia Zawadzka}
\affiliation{\FUW}
\author{Dipankar Jana}
\affiliation{\Grenoble}
\author{Igor Antoniazzi}
\affiliation{\FUW}
\author{Magdalena~Grzeszczyk}
\affiliation{\Singapore}
\author{Ma\l{}gorzata Zinkiewicz}
\affiliation{\FUW}
\author{Kenji~Watanabe}
\affiliation{\Watanabe}
\author{Takashi Taniguchi}
\affiliation{\Taniguchi}
\author{Marek~Potemski}
\affiliation{\FUW}
\affiliation{\Grenoble}
\affiliation{\Centera}
\author{Cl\'ement Faugeras}
\affiliation{\Grenoble}
\author{Adam Babi\'nski}
\affiliation{\FUW}
\author{Maciej R. Molas}
\email{maciej.molas@fuw.edu.pl}
\affiliation{\FUW}
\maketitle

\section{Low-temperature PL spectrum of WSe$_2$ monolayer \label{sec:S1}}

\begin{figure}[h]
		\subfloat{}%
		\centering
		\includegraphics[width=0.7 \linewidth]{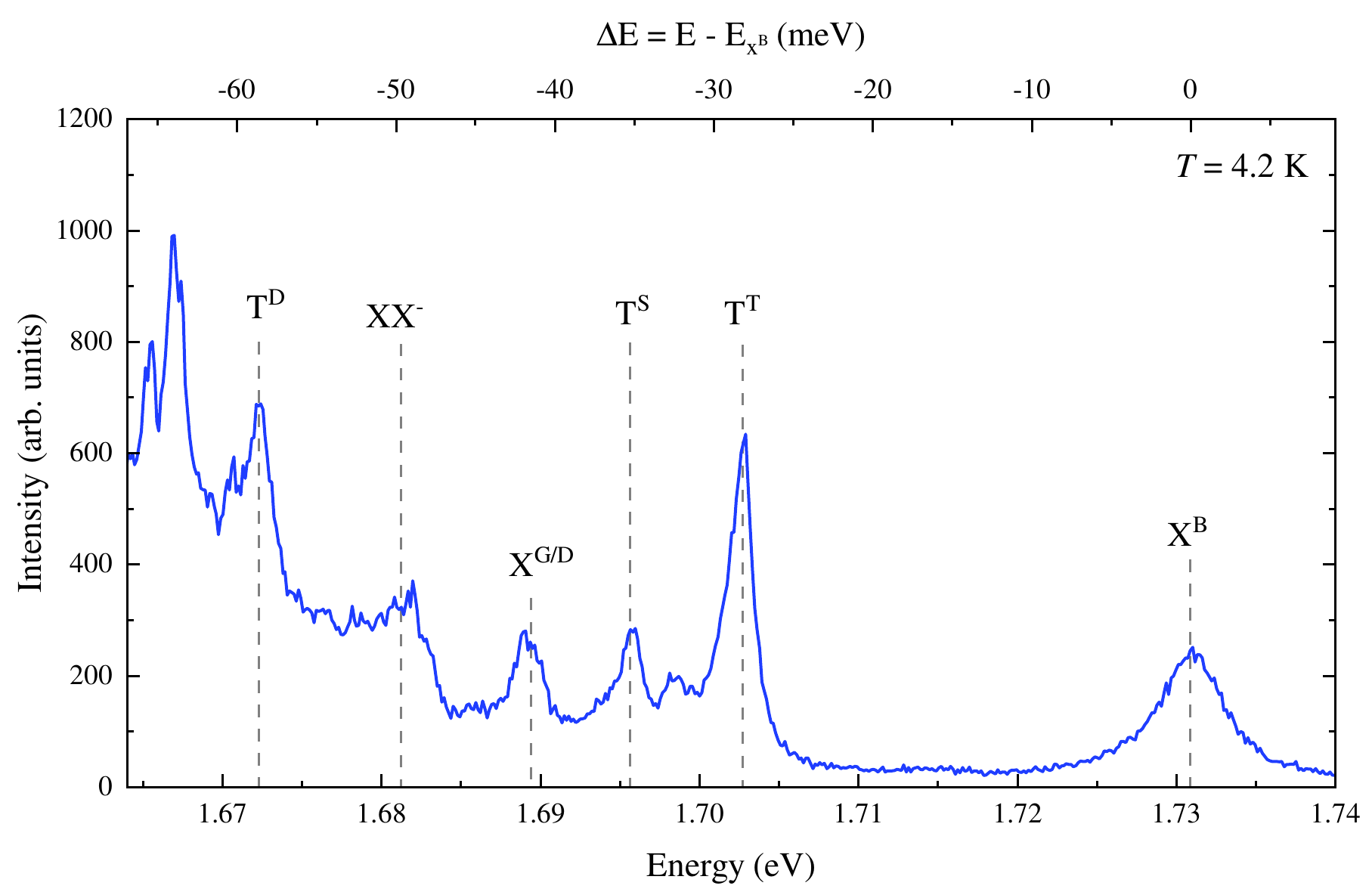}
    	\caption{Low-temperature ($T$=4.2~K) PL spectrum of the WSe$_2$ ML encapsulated in hBN flakes measured using excitation energy 2.41~eV and power of 10~$\mu$eV. 
        Lines assignments is a follows: X$^\textrm{B}$ -- neutral exciton; 
        T$^\textrm{S}$ and T$^\textrm{T}$ -- singlet (intravalley) and triplet (intervalley) negatively charged excitons, respectively; 
        X$^\textrm{G}$ -- gray exciton, 
        XX$^-$ -- negatively charged biexciton; 
        T$^\textrm{D}$ -- negatively charged dark exciton (dark trion).}
		\label{fig_WSe2}
\end{figure}

Fig.~\ref{fig_WSe2} shows the low-temperature ($T$=4.2~K) PL spectrum measured on a WSe$_2$ ML encapsulated in hexagonal BN (hBN) flakes.
The spectrum displays several emission lines with a characteristic pattern similar to that previously reported in several works on WSe$_2$ MLs embedded in between hBN flakes~\cite{Courtade2017, Li2018, Chen2018, Barbone2018, Paur2019, Liu2019, Li2019, Li2019replica, Li2019momentum, Molas2019, LiuValley, Liu2020, He2020, Robert2021, Robert2021PRL, Arora2015, Smolenski2016, Robert2017, Wang2017, Koperski2017, Koperski2019, Arora2020, Zinkiewicz2022}. 
According to these reports, the assignment of the observed emission lines is as follows:
X$^\textrm{B}$ -- neutral exciton;  
T$^\textrm{S}$ and T$^\textrm{T}$ -- singlet (intravalley) and triplet (intervalley) negatively charged excitons, respectively; 
X$^\textrm{G}$ -- grey exciton;
XX$^-$ -- negatively charged biexciton; 
T$^\textrm{D}$ -- negatively charged dark exciton (dark trion);

\section{Magnetic field evolution of the integrated intensities of the dark excitons and trion in different temperatures \label{sec:S2}}

\begin{figure}[h!]
		\subfloat{}%
		\centering
		\includegraphics[width=0.94 \linewidth]{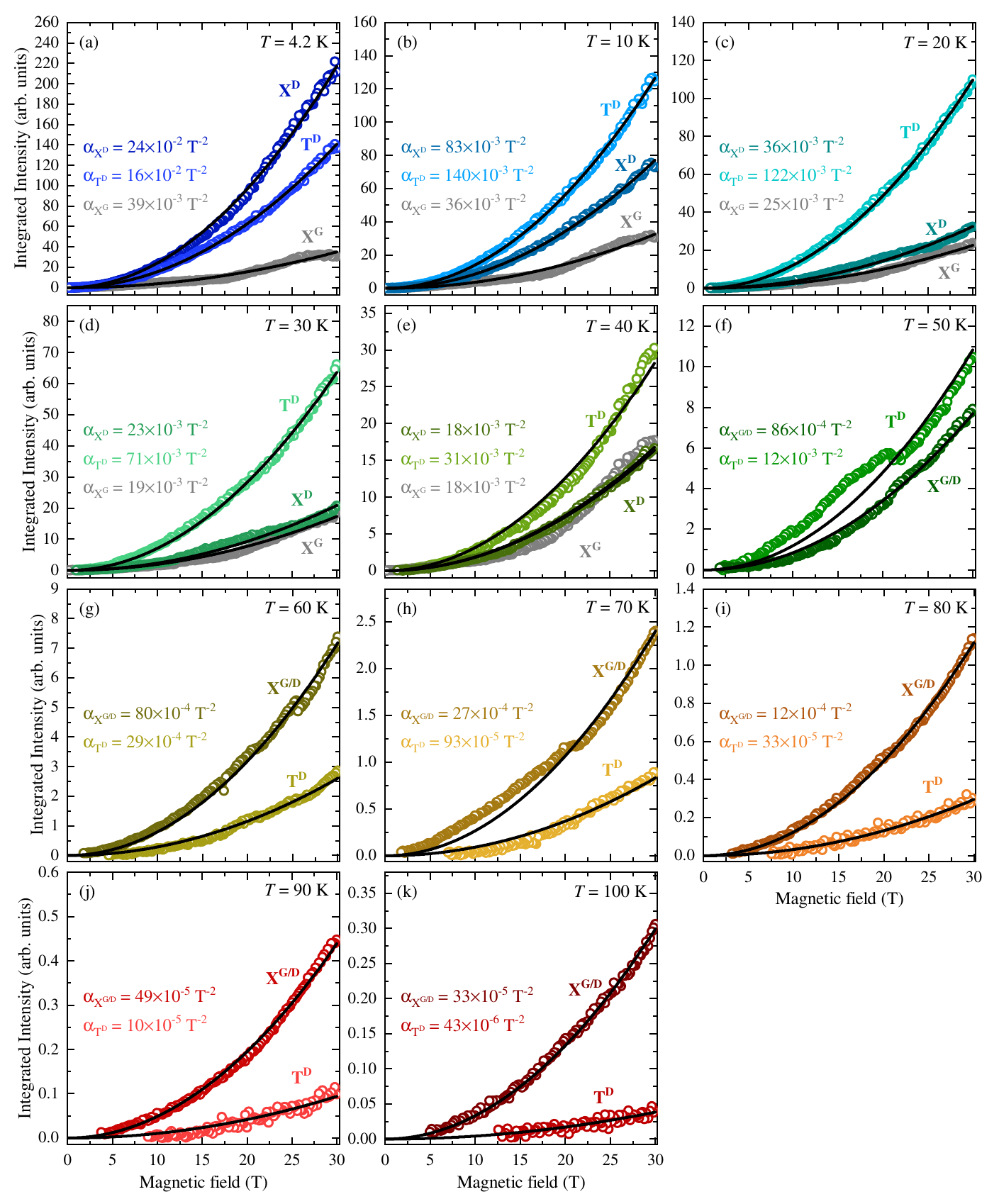}
    	\caption{Magnetic field evolution of the integrated intensities of the neutral grey (X$^\textrm{G}$) and dark (X$^\textrm{D}$) excitons and the dark trion (T$^\textrm{D}$) extracted at different temperatures. Black curves represent the fits of the function $I = \alpha B_\parallel^{2}$.}
		\label{fig_s2}
\end{figure}

The evolutions of the integrated intensities of the neutral grey (X$^\textrm{G}$) and dark (X$^\textrm{D}$) excitons and the dark trion (T$^\textrm{D}$) as a function of the external in-plane magnetic field ($B_\parallel$) extracted at different temperatures from 4.2~K to 100~K are presented in Fig.~\ref{fig_s2}. 
The brightening of these dark emissions follows the quadratic evolution of their integrated intensities ($I$) as a function of $B_\parallel$, $i.e.$, $I=\alpha B_\parallel^{2}$, where $\alpha$ is a fitting parameter.

\section{Temperature evolution of the $\alpha$ parameter of the dark excitons and trion \label{sec:S3}}

\begin{figure}[h!]
		\subfloat{}%
		\centering
		\includegraphics[width=0.45 \linewidth]{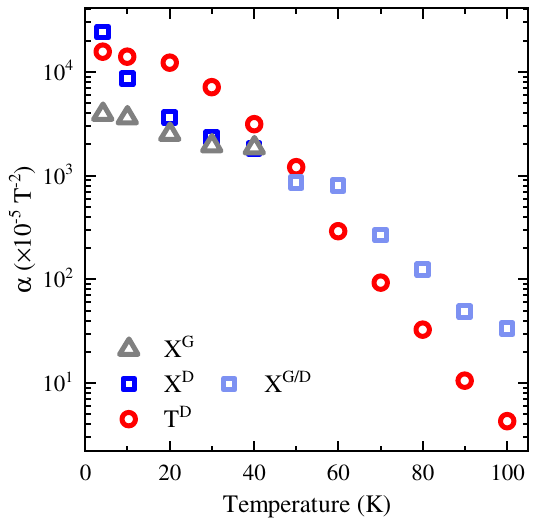}
    	\caption{Temperature evolution of the $\alpha$ parameter of the neutral grey (X$^\textrm{G}$) and dark (X$^\textrm{D}$) excitons and the dark trion (T$^\textrm{D}$) extracted by fitting of the function $I = \alpha B_\parallel^{2}$.}
		\label{fig_s3}
\end{figure}

Figure~\ref{fig_s3} shows the temperature dependences of the $\alpha$ parameters of the T$^\textrm{D}$, X$^\textrm{D}$, and X$^\textrm{G}$ complexes.
The evolutions of the $\alpha$ parameters as a function of temperature are analogous to the ones of their extracted integrated intensities of the integrated intensities presented in the main articles.

\section{Temperature evolutions of the energies of the bright and dark excitons and trions \label{sec:S4}}

\begin{figure}[h!]
		\subfloat{}%
		\centering
		\includegraphics[width=0.75 \linewidth]{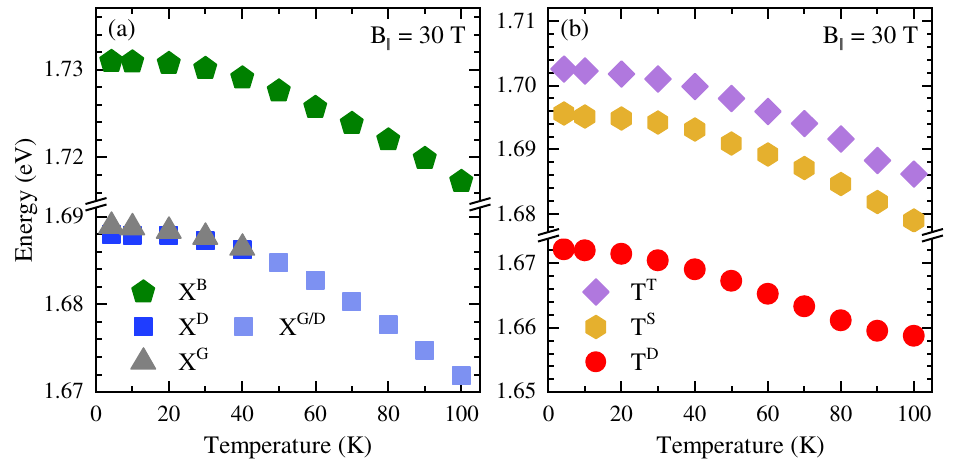}
    	\caption{Energy evolutions of the (a) neutral excitons: bright (X$^\textrm{B}$), grey (X$^\textrm{G}$) dark (X$^\textrm{D}$), and the (b) negative trions:  spin-singlet (T$^\textrm{S}$), spin-triplet (T$^\textrm{T}$), and dark (T$^\textrm{D}$).}
		\label{fig_s4}
\end{figure}

The temperature dependences of the neutral excitons: bright (X$^\textrm{B}$), grey (X$^\textrm{G}$) dark (X$^\textrm{D}$), and the negative trions: spin-singlet (T$^\textrm{S}$), spin-triplet (T$^\textrm{T}$), and dark (T$^\textrm{D}$) are presented in Fig.~\ref{fig_s4}.
For all traces shown in Fig.~\ref{fig_s4}, the energies of all excitonic resonances experience a redshift as the temperature increases from 4.2 K to 100 K. 
This type of evolution is characteristic of many semiconductors, particularly for intralayer transitions in thin layers of S-TMD~\cite{Arora2015, Molas2017nano}.

\newpage
\bibliographystyle{apsrev4-2}
\bibliography{biblio}